\documentclass{ifacconf}

\makeatletter
\let\old@ssect\@ssect
\makeatother

\usepackage[pdfencoding=auto,psdextra,hidelinks]{hyperref}
\usepackage{natbib}
\usepackage{lmodern}
\usepackage{amsmath}
\usepackage{amssymb}
\usepackage{bm}
\usepackage{graphicx}
\usepackage{xcolor}
\usepackage{booktabs}
\usepackage[font=footnotesize]{subfig}
\usepackage{eso-pic}
\usepackage{etoolbox}
\usepackage{textcomp}

\newcommand{\D}{\displaystyle}

\newcommand{\Vector}[1]{\bm{#1}}  
\newcommand{\Matrix}[1]{\bm{#1}}  
\newcommand{\Transpose}{\mathrm{T}}  

\newcommand{\refEq}[1]{(\ref{#1})}               
\newcommand{\refFig}[1]{Figure~\ref{#1}}         
\newcommand{\refSec}[1]{Section~\ref{#1}}        

\graphicspath{{./gfx/}}

\makeatletter
\def\@ssect#1#2#3#4#5#6{%
  \NR@gettitle{#6}
  \old@ssect{#1}{#2}{#3}{#4}{#5}{#6}
}
\makeatother

\hypersetup{%
    pdfauthor={Gernot Herbst, Arne-Jens Hempel, Thomas G\"{o}hrt, Stefan Streif},
    pdftitle={Half-Gain Tuning for Active Disturbance Rejection Control},
    pdfcreator={},pdfproducer={}
}

\begin{document}

\AddToShipoutPicture*{\AtPageUpperLeft{\put(\LenToUnit{1.5cm},\LenToUnit{-2.5cm}){%
   \color{black!50}%
   \begin{minipage}[t]{18.5cm}
   \footnotesize%
   \textcopyright\ 2020 the authors. This work has been accepted to IFAC WC 2020 for publication under a Creative Commons License CC BY-NC-ND.\\%
   The final article is available at \textcolor{blue!60}{\href{https://doi.org/10.1016/j.ifacol.2020.12.1864}{https://doi.org/10.1016/j.ifacol.2020.12.1864}}.%
   \end{minipage}
}}}

\begin{frontmatter}

\title{Half-Gain Tuning for\\ Active Disturbance Rejection Control}

\author[First]{Gernot Herbst}
\author[Second]{Arne-Jens Hempel}
\author[Second]{Thomas G\"{o}hrt}
\author[Second]{Stefan Streif}

\address[First]{Siemens AG, Clemens-Winkler-Str. 3, 09116 Chemnitz, Germany}
\address[Second]{Technische Universit\"{a}t Chemnitz, 09107 Chemnitz, Germany}

\begin{abstract}  
A new tuning rule is introduced for linear active disturbance rejection control (ADRC), which results in similar closed-loop dynamics as the commonly employed bandwidth parameterization design, but with lower feedback gains.
In this manner the noise sensitivity of the controller is reduced, paving the way for using ADRC in more noise-affected applications.
It is proved that the proposed tuning gains, while rooted in the analytical solution of an algebraic Riccati equation, can always be obtained from a bandwidth parameterization design by simply halving the gains. This establishes a link between optimal control and pole placement design.
\end{abstract}


\end{frontmatter}


\section{Introduction}

ADRC was developed as a nonlinear general-purpose controller by \cite{Han:2009}.
A linear variant was proposed by \cite{Gao:2003}, facilitating stability analysis and significantly reducing the number of tuning parameters with the introduction of the ``bandwidth parameterization'' approach.

The seemingly unorthodox use of elements from modern control theory for creating an almost model-free controller from the user's point of view is key to its appraisal as a ``paradigm change'', cf.\ \cite{Gao:2001,Gao:2006}, and a differentiator to other model-free approaches, e.\,g.\ as introduced by \cite{Fliess:2013}.
And indeed, the ease of tuning, its performance compared to traditional (PID-type) control, and its extendability with features desirable for industrial applications as in \cite{Herbst:2016a} and \cite{Madonski:2019}, make it an attractive alternative for real-world control problems, cf.\ \cite{Zheng:2018}.

The core of ADRC is formed by an observer, which is denoted as ``extended state observer'' (ESO) and puts emphasis on rejecting disturbances in a broader sense.
There are further possible extensions to the observer, such as tracking disturbance derivatives using Generalized Proportional Integral (GPI) observers, cf.\ \cite{Sira:2017}. However, we will focus on the (arguably) most common variant of the ESO, which incorporates a single lumped (``total'') disturbance state modeling both unknown dynamics and piecewise constant input disturbance signals of the plant.

In this paper, we will explore the use of the so-called $\alpha$-controller design for tuning the observer and control loop within linear ADRC.
It was put forward by \cite{Buhl:2009} and is based on the solution of an algebraic Riccati equation to obtain feedback gains leading to an exponentially decaying Lyapunov function for the controlled system.
A similar approach has been proposed by \cite{Zhou:2008a}, denoted as ``low gain feedback''.
As a matter of fact, applying $\alpha$-controller design to ADRC will lead to reduced controller/observer gains compared to the established bandwidth parameterization approach, which will in turn reduce noise sensitivity of the resulting design.

The main contribution of this work is the introduction of a new ADRC tuning rule, which we will denote as ``half-gain tuning''.
We will show that $\alpha$-controller design aiming at closed-loop dynamics similar to bandwidth parameterization will always lead to exactly halved gains for the controller and/or observer.
Therefore, while grounded in an algebraic Riccati equation, an $\alpha$-controller design for ADRC can be trivially obtained from bandwidth parameterization, superseding the need for solving the former.
For an example, detailed insights are given into the frequency- and time-domain behavior when using ADRC with half-gain tuning for the controller and/or observer.


\section{Active Disturbance Rejection Control}
\label{sec:ADRC}

This section provides a very brief overview of continuous-time linear ADRC and the prevalent tuning method.
For a more detailed introduction we refer to \cite{Herbst:2013}.


\subsection{Idea and Structure of the Controller}
\label{sec:ADRC_approach}

The essence of linear ADRC can be described as follows:
\begin{enumerate}
\item
assume an $n$-th order integrator chain behavior $P(s) = b_0/s^n$ for a single-input single-output (SISO) plant of order $n$, regardless of its actual structure;
\item
apply the inverted gain $1/b_0$ at the controller output to compensate for the plant gain $b_0$;
\item
set up a full-order observer for the the integrator chain model (estimated states $\hat{x}_{1,\ldots,n}$), extended by a constant input disturbance (estimate $\hat{x}_{n+1}$) that captures both actual disturbances and model uncertainties (``extended state observer'', ESO);
\item
compensate the disturbance using the estimate $\hat{x}_{n+1}$;
\item
design a full-order state-feedback controller for the remaining ``pure'' integrator chain $1/s^n$ to achieve the desired closed-loop dynamics.
\end{enumerate}

\begin{figure}[ht]
    \centering%
    \includegraphics[width=\linewidth]{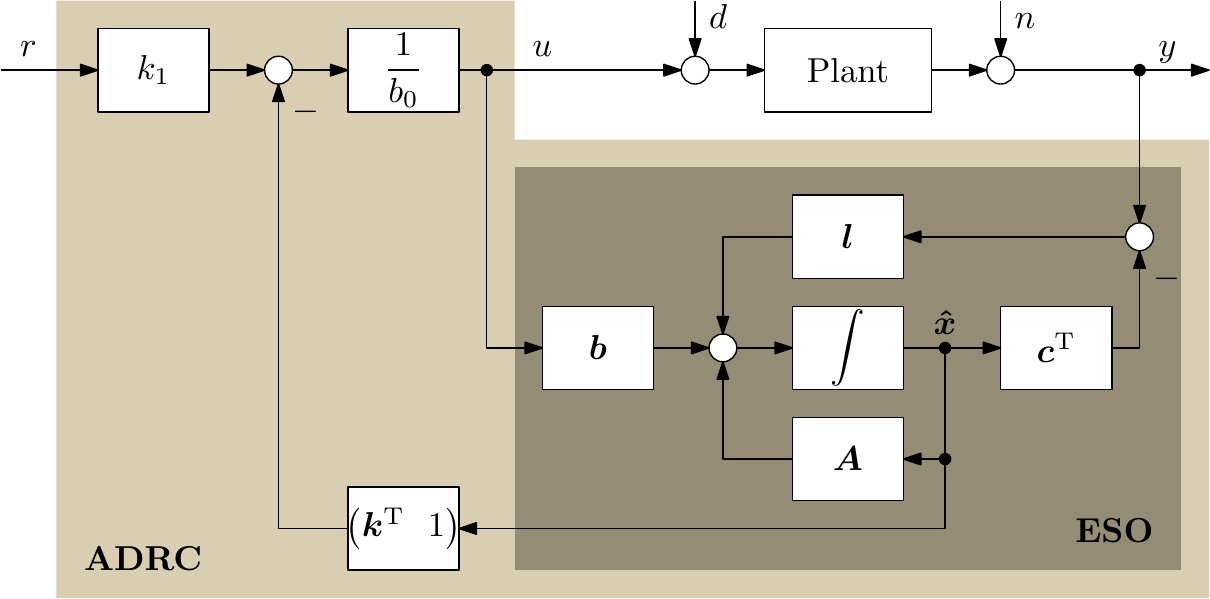}%
    \caption{Control loop with ADRC, consisting of extended state observer (ESO) and state-feedback controller including disturbance compensation. Signals: controlled variable $y$, control action $u$, reference $r$; and possible disturbances at plant input ($d$) and output ($n$).}
    \label{fig:ADRC_ControlLoop}
\end{figure}

Control law and observer equations are illustrated in \refFig{fig:ADRC_ControlLoop}, and given in \refEq{eqn:ADRC_Controller} and \refEq{eqn:ADRC_Observer} for the $n$-th order case.
\begin{gather}
u(t) = \frac{1}{b_0} \cdot \left( k_1 \cdot r(t) -
\begin{pmatrix}
\Vector{k}^\Transpose  &  1
\end{pmatrix}
\cdot \Vector{\hat{x}}(t)
\right)
\label{eqn:ADRC_Controller}
\\
\text{with}\quad
\Vector{k}^\Transpose =
\begin{pmatrix}
k_1  &  \cdots  &  k_n
\end{pmatrix}
,\quad
\Vector{\hat{x}} =
\begin{pmatrix}
\hat{x}_1  &  \cdots  &  \hat{x}_{n+1}
\end{pmatrix}^\Transpose
\notag
\end{gather}

\begin{gather}
\Vector{\dot{\hat{x}}}(t)
=
\Matrix{A}_\mathrm{ESO} \cdot \Vector{\hat{x}}(t)
+
\Vector{b}_\mathrm{ESO} \cdot u(t)
+
\Vector{l} \cdot \left( y(t) - \Vector{c}_\mathrm{ESO}^\Transpose \cdot  \Vector{\hat{x}}(t) \right)
\label{eqn:ADRC_Observer}
\\
\text{with}\quad
\Matrix{A}_\mathrm{ESO} =
\begin{pmatrix}
\Matrix{0}^{n \times 1}  &  \Matrix{I}^{n \times n}
\\
0  &  \Matrix{0}^{1 \times n}
\end{pmatrix}
,\quad
\Vector{b}_\mathrm{ESO} =
\begin{pmatrix}
\Matrix{0}^{(n-1) \times 1}
\\
b_0
\\
0
\end{pmatrix}
,\notag\\
\Vector{l} =
\begin{pmatrix}
l_1  &  \cdots  &  l_{n+1}
\end{pmatrix}^\Transpose
,\quad
\Matrix{c}_\mathrm{ESO}^\Transpose =
\begin{pmatrix}
1  &  \Matrix{0}^{1 \times n}
\end{pmatrix}
\notag
\end{gather}

\subsection{Tuning by Bandwidth Parameterization}
\label{sec:ADRC_tuning}

Assuming perfect disturbance rejection and compensation of the plant gain $b_0$ by the control law \refEq{eqn:ADRC_Controller}, the state-feedback controller $\Vector{k}^\Transpose$ has to be designed for a ``virtual'' plant in form of a pure $n$-th order integrator chain:
\begin{gather}
\Vector{\dot{x}}(t)
=
\Matrix{A} \cdot \Vector{x}(t)
+
\Vector{b} \cdot u(t)
,\quad
y(t) = x_1(t)
\label{eqn:CtrlDesign_Plant}
\\
\text{with}\quad
\Matrix{A} =
\begin{pmatrix}
\Matrix{0}^{(n-1) \times 1}  &  \Matrix{I}^{(n-1) \times (n-1)}
\\
0  &  \Matrix{0}^{1 \times (n-1)}
\end{pmatrix}
,\quad
\Vector{b} =
\begin{pmatrix}
\Matrix{0}^{(n-1) \times 1}
\\
1
\end{pmatrix}
.
\notag
\end{gather}

The predominant controller design approach in linear ADRC is called ``bandwidth parameterization'', cf.\ \cite{Gao:2003}, and is using pole placement with all poles at $\lambda = -\omega_\mathrm{CL}$, the desired closed-loop bandwidth:
\begin{align}
\left( \lambda + \omega_\mathrm{CL} \right)^n
& \stackrel{!}{=}
\det\left( \lambda \Matrix{I} - \left( \Matrix{A} - \Vector{b} \Vector{k}^\Transpose \right) \right)
\label{eqn:CtrlDesign_K_PolePlacement}
\\
&= \lambda^{n} + k_n \lambda^{n-1} + \ldots + k_2 \lambda + k_1
.
\notag
\end{align}

Binominal expansion of \refEq{eqn:CtrlDesign_K_PolePlacement} leads to the controller gains:
\begin{equation}
k_i = \frac{n!}{(n-i+1)! \cdot (i-1)!} \cdot \omega_\mathrm{CL}^{n-i+1}
\quad \forall i = 1, ..., n
.
\label{eqn:CtrlDesign_K_PolePlacement_Param}
\end{equation}

For tuning the extended state observer (ESO) with the bandwidth approach, we will follow the notation of \cite{Herbst:2013} in placing the closed-loop observer poles at $\lambda = -k_\mathrm{ESO} \cdot \omega_\mathrm{CL}$, with $k_\mathrm{ESO}$ being the relative factor between observer and control loop bandwidth:
\begin{align}
\left( \lambda + k_\mathrm{ESO} \cdot \omega_\mathrm{CL} \right)^{n+1}
& \stackrel{!}{=}
\det\left( \lambda \Matrix{I} - \left( \Matrix{A}_\mathrm{ESO} - \Vector{l} \Vector{c}_\mathrm{ESO}^\Transpose \right) \right)
\label{eqn:ObsDesign_L_PolePlacement}
\\
&= \lambda^{n+1} + l_1 \lambda^{n} + \ldots + l_{n} \lambda + l_{n+1}
.
\notag
\end{align}

Binominal expansion of \refEq{eqn:ObsDesign_L_PolePlacement} yields the observer gains:
\begin{equation}
l_i = \frac{(n+1)!}{(n+1-i)! \cdot i!} \cdot \left( k_\mathrm{ESO} \cdot \omega_\mathrm{CL} \right)^i
\quad \forall i = 1, ..., n+1
.
\label{eqn:ObsDesign_L_PolePlacement_Param}
\end{equation}

Comparing these two tuning tasks for linear ADRC, we can conclude that---in both cases---only integrator chains are to be handled: of order $n$ (for the closed-loop dynamics) and $n+1$ (for the extended state observer).


\section{\texorpdfstring{$\alpha$}{\textalpha}-Controller Approach}
\label{sec:alpha}


\subsection{Brief Overview of the Tuning Method}
\label{sec:alpha_intro}

\cite{Buhl:2009} introduced the so-called $\alpha$-controller approach, a design method leading to an exponentially decreasing Lyapunov function for the closed-loop system. The rate of decay $\alpha$ is the only design parameter of this method:
\begin{equation}
\dot{V} = -\alpha V
\quad\text{with}\quad
\alpha > 0
\quad\text{and}\quad
V = \Vector{x}^\Transpose \Matrix{P} \Vector{x}
.
\label{eqn:Alpha_Lyapunov}
\end{equation}

With a plant as in \refEq{eqn:CtrlDesign_Plant}:
\begin{equation}
\begin{split}
\dot{V} &= \left( \frac{\partial V}{\partial \Vector{x}} \right)^\Transpose \Vector{\dot{x}}
= 2 \Vector{x}^\Transpose \Matrix{P} \left( \Matrix{A} \Vector{x} + \Vector{b} u \right)
\\
&= \Vector{x}^\Transpose \left( \Matrix{A}^\Transpose \Matrix{P} + \Matrix{P} \Matrix{A} \right) \Vector{x} + 2 \Vector{x}^\Transpose \Matrix{P} \Vector{b} u
\\
&= - \alpha \Vector{x}^\Transpose \Matrix{P} \Vector{x}
.
\end{split}
\label{eqn:Alpha_Derive}
\end{equation}

A suitable control law for achieving a negative $\dot{V}$ in \refEq{eqn:Alpha_Derive} is:
\begin{equation}
u = -\Vector{b}^\Transpose \Matrix{P} \Vector{x}
.
\label{eqn:Alpha_ControlLaw}
\end{equation}

Combining these two equations we obtain an algebraic Riccati equation:
\begin{equation}
\left( \Matrix{A} + \frac{\alpha}{2} \Matrix{I} \right)^\Transpose \Matrix{P}
+ \Matrix{P} \left( \Matrix{A} + \frac{\alpha}{2} \Matrix{I} \right)
- 2 \Matrix{P} \Vector{b} \Vector{b}^\Transpose \Matrix{P}
= 0
.
\label{eqn:Alpha_Riccati}
\end{equation}

In summary, the state-feedback controller gains for the $\alpha$-controller approach are $\Vector{k}^\Transpose = \Vector{b}^\Transpose \Matrix{P}$, with $\Matrix{P}$ being the solution of \refEq{eqn:Alpha_Riccati}.


\subsection{Comparison to Bandwidth Parameterization}
\label{sec:alpha_comparison}

When applying the $\alpha$-controller tuning approach to a loop with the plant being an integrator chain, the closed-loop poles may be complex-valued, but will all have the same real part of $-\frac{\alpha}{2}$, cf.\ the proof in \cite{Buhl:2009}. On the other hand, using the ``bandwidth parameterization'' pole placement design as given in \refEq{eqn:CtrlDesign_K_PolePlacement}, all closed-loop poles will be at $-\omega_\mathrm{CL}$ and real-valued only.

\begin{figure}[ht]
    \centering%
    \subfloat[With plant being integrator chain of order $n = 2$]{%
        \includegraphics{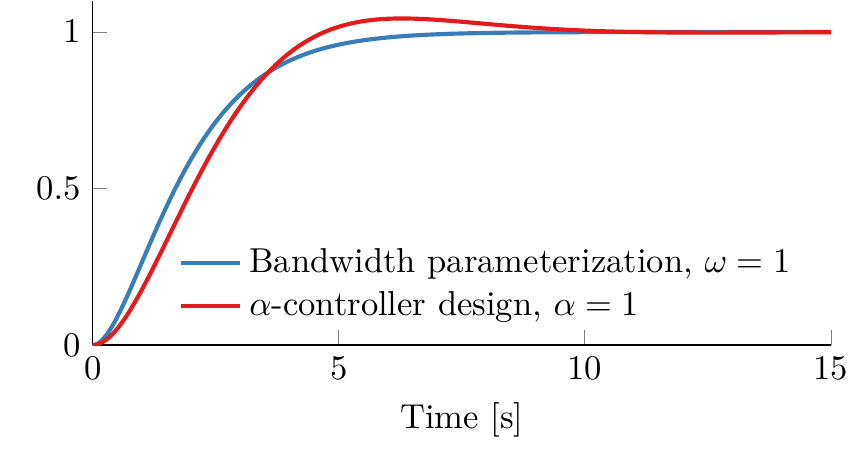}%
    }%
    \\[0.3cm]
    \subfloat[With plant being integrator chain of order $n = 3$]{%
        \includegraphics{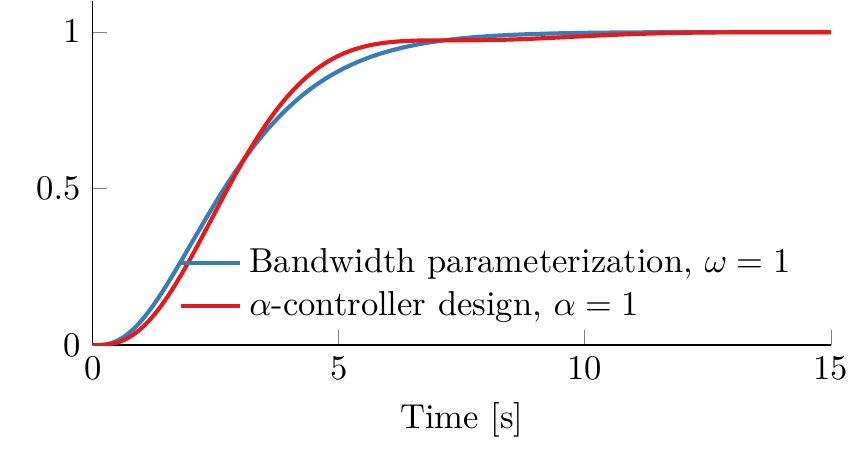}%
    }%
    \caption{Comparison of the normalized closed-loop step responses using bandwidth parameterization (pole placement) and $\alpha$-controller design for full-order state-feedback control of integrator chain systems of order $n = 2$ and $n = 3$.}
    \label{fig:Comparison_N2_N3}
\end{figure}

Being the respective counterparts of PI and PID controllers, first- and second-order ADRC are the most relevant cases in practical applications, resulting in tuning tasks for integrator chains of order up to three. When comparing the $\alpha$-controller tuning results with pole placement (bandwidth parameterization for $\omega_\mathrm{CL}$), two observations can be made:
\begin{enumerate}
\item
Selecting the tuning parameters of both methods as $\alpha = \omega_\mathrm{CL}$ results in similar closed-loop dynamics for integrator chains of order two and above, with a slightly underdamped response for the $\alpha$-controller due to the complex-valued poles. For the second- and third-order case, the closed-loop step response achieved with these two methods is being compared in \refFig{fig:Comparison_N2_N3}. A first-order $\alpha$-controller design would be necessarily slower, since only one real-valued pole (at $-\frac{\alpha}{2}$, therefore at half the bandwidth of $\omega_\mathrm{CL}$) can be placed.

\item
When designing with $\alpha = \omega_\mathrm{CL}$, the resulting controller gains of the $\alpha$-controller approach are exactly half of the controller gains obtained using pole placement with bandwidth parameterization. We will prove this relation in \refSec{sec:halfgain}.
\end{enumerate}

\begin{figure}[ht]
    \centering%
    \subfloat[$n = 2$]{%
        \includegraphics{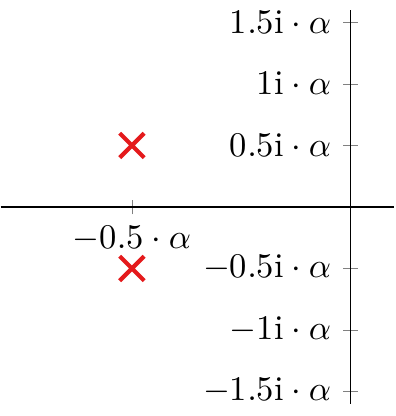}%
    }%
    \hfill%
    \subfloat[$n = 3$]{%
        \includegraphics{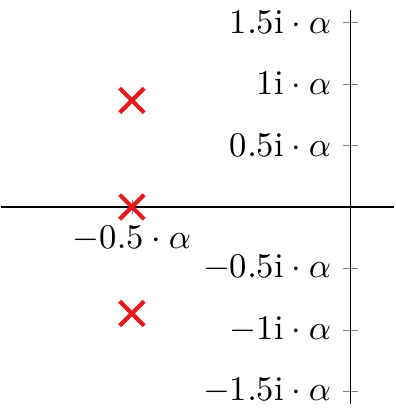}%
    }%
    \\
    \subfloat[$n = 4$]{%
        \includegraphics{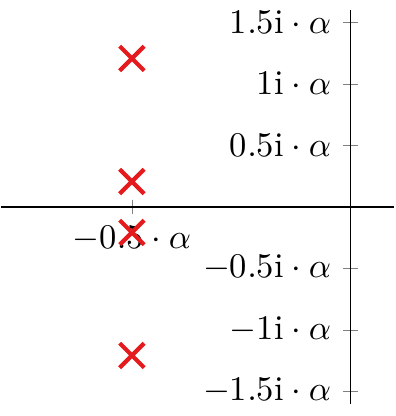}%
    }%
    \hfill%
    \subfloat[$n = 5$]{%
        \includegraphics{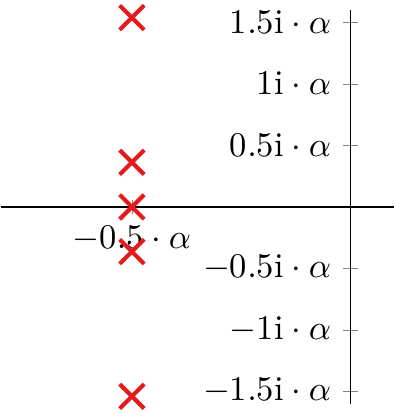}%
    }%
    \caption{Closed-loop poles resulting from $\alpha$-controller design for integrator chain plants of orders $n = 2$ to $n = 5$. Bandwidth parameterization, by contrast, will always place all poles at $-\alpha$.}
    \label{fig:Comparison_Alpha_Poles}
\end{figure}

The closed-loop pole configurations of $\alpha$-controller designs are presented in \refFig{fig:Comparison_Alpha_Poles}, with the most important cases being:
\begin{itemize}
\item
$\D \lambda_{\mathrm{1/2}} = \left( -\frac{1}{2} \pm \frac{1}{2} \mathrm{i} \right) \cdot \alpha$
\quad
for the second-order integrator chain, and

\item
$\D \lambda_\mathrm{1} = -\frac{1}{2} \cdot \alpha, \quad \lambda_{\mathrm{2/3}} = \left( -\frac{1}{2} \pm \frac{\sqrt{3}}{2} \mathrm{i} \right) \cdot \alpha$
\quad
for the third-order integrator chain.
\end{itemize}

Concluding this comparison: The $\alpha$-controller design leads to similar closed-loop dynamics for systems of order two and above, but with only half the controller gain compared to a pole placement design with bandwidth parameterization. Therefore the impact of measurement noise on the control action will be reduced, making the $\alpha$-controller design an interesting alternative for noise-affected systems, if the underdamped behavior is tolerable.


\section{The Half-Gain Tuning Method for ADRC}
\label{sec:halfgain}


As pointed out in \refSec{sec:alpha_comparison}, there are up to three options of replacing bandwidth parameterization in linear ADRC with the $\alpha$-controller approach: (1) only for the controller (using $\alpha = \omega_\mathrm{CL}$, for ADRC of order $n \ge 2$); (2) only for the observer (using $\alpha = k_\mathrm{ESO} \cdot \omega_\mathrm{CL}$, possible for any order $n \ge 1$); or (3) for both controller and observer (for $n \ge 2$).

Applying $\alpha$-controller tuning to ADRC results in halved controller and/or observer gains, while maintaining similar (albeit slightly underdamped) dynamics for the control loop and/or the extended state observer. We will therefore denote this design approach for ADRC as the ``half-gain tuning'' method.

This is the main result of this article, which will be proved in the following:
To obtain the $\alpha$-controller gains, the Riccati equation \refEq{eqn:Alpha_Riccati} does \emph{not} have to be solved. The gains can simply be obtained from the straightforward bandwidth tuning rules, i.\,e.\ equations \refEq{eqn:CtrlDesign_K_PolePlacement_Param} (controller) or \refEq{eqn:ObsDesign_L_PolePlacement_Param} (observer), by halving these gains for a bandwidth $\omega_\mathrm{CL} = \alpha$ (controller) or $k_\mathrm{ESO} \cdot \omega_\mathrm{CL} = \alpha$ (observer).

\begin{thm}
For plants as given in \refEq{eqn:CtrlDesign_Plant}, the controller gains $\Vector{k}^\Transpose_\mathrm{BW}$ obtained via bandwidth parameterization in \refEq{eqn:CtrlDesign_K_PolePlacement} are related to the $\alpha$-controller gains $\Vector{k}^\Transpose_\alpha$ from \refEq{eqn:Alpha_ControlLaw}, \refEq{eqn:Alpha_Riccati} by an exact factor of two, if $\Vector{k}^\Transpose_\mathrm{BW}$ has been designed for a bandwidth $\omega_\mathrm{CL} = \alpha$:
\begin{equation}
\Vector{k}^\Transpose_\alpha = \frac{1}{2} \cdot \Vector{k}^\Transpose_\mathrm{BW}
= \frac{1}{2} \cdot \begin{pmatrix}
k_{\mathrm{BW},1}  &  \cdots  &  k_{\mathrm{BW},n}
\end{pmatrix}
.
\label{eqn:K_Relation}
\end{equation}
\end{thm}

\begin{pf}
We start by rewriting \refEq{eqn:Alpha_Riccati} as follows,
\begin{equation}
\alpha \Matrix{P} = -\left( \Matrix{A}^\Transpose \Matrix{P} + \Matrix{P} \Matrix{A} \right) + \Matrix{S},
\label{eqn:Proof_1}
\end{equation}
where, using \refEq{eqn:Alpha_ControlLaw} and \refEq{eqn:K_Relation},
\begin{equation}
\Matrix{S} = \frac{1}{2} \Vector{k}_\mathrm{BW} \Vector{k}^\Transpose_\mathrm{BW}
= \frac{1}{2} \cdot \begin{pmatrix}
k_{\mathrm{BW},1} \Vector{k}_\mathrm{BW}  &  \cdots  &  k_{\mathrm{BW},n} \Vector{k}_\mathrm{BW}
\end{pmatrix}
.
\end{equation}

Since $\Matrix{A}$ is an upper shift matrix, $\Matrix{P} \Matrix{A}$ will result in $\Matrix{P}$'s columns $\Vector{p}_i$ being shifted:
\begin{equation}
\Matrix{P} \Matrix{A}
=
\begin{pmatrix}
\Vector{0}  &  \Vector{p}_1  &  \cdots  &  \Vector{p}_{n-1}
\end{pmatrix}
.
\end{equation}

From the first column of \refEq{eqn:Proof_1} we obtain $\Vector{p}_1$, and, as an abbreviation, introduce $\Matrix{\Phi}$:
\begin{align}
\alpha \Vector{p}_1 &= -\Matrix{A}^\Transpose \Vector{p}_1 + \frac{k_{\mathrm{BW},1}}{2} \Vector{k}_\mathrm{BW}
\notag\\
\Vector{p}_1 &= \left( \alpha \Matrix{I} + \Matrix{A}^\Transpose \right)^{-1} \cdot \frac{k_{\mathrm{BW},1}}{2} \Vector{k}_\mathrm{BW}
= \Matrix{\Phi}^{-1} \cdot \frac{k_{\mathrm{BW},1}}{2} \Vector{k}_\mathrm{BW}
.
\end{align}

For all other columns ($i = 2, \ldots, n$):
\begin{align}
\alpha \Vector{p}_i &= -\Matrix{A}^\Transpose \Vector{p}_i - \Vector{p}_{i-1} + \frac{k_{\mathrm{BW},i}}{2} \Vector{k}_\mathrm{BW}
\notag\\
\Vector{p}_i &= -\Matrix{\Phi}^{-1} \cdot \Vector{p}_{i-1} + \Matrix{\Phi}^{-1} \cdot \frac{k_{\mathrm{BW},i}}{2} \Vector{k}_\mathrm{BW}
.
\label{eqn:Proof_2}
\end{align}

We now recursively expand \refEq{eqn:Proof_2} for the final ($n$-th) column:
\begin{equation}
\Vector{p}_n
= \sum_{i = 1}^n (-1)^{(n-i)} \cdot \Matrix{\Phi}^{-(n-i+1)} \cdot \frac{k_{\mathrm{BW},i}}{2} \Vector{k}_\mathrm{BW}
.
\label{eqn:Proof_pn}
\end{equation}

$\Vector{p}_n^\Transpose$ is the gain vector of the $\alpha$-controller, since, recalling \refEq{eqn:Alpha_ControlLaw} with \refEq{eqn:CtrlDesign_Plant}, $\Vector{k}_\alpha^\Transpose = \Vector{b}^\Transpose \Matrix{P} = \Vector{b}^\Transpose \Matrix{P}^\Transpose = \Vector{p}_n^\Transpose$.
Multiplying \refEq{eqn:Proof_pn} with $\Matrix{\Phi}^n$ one obtains:
\begin{equation}
\Matrix{\Phi}^n \cdot \Vector{p}_n
= \left( \sum_{i = 1}^n (-1)^{(n-i)} \cdot \Matrix{\Phi}^{(i-1)} \cdot k_{\mathrm{BW},i} \right) \cdot \frac{1}{2} \Vector{k}_\mathrm{BW}
.
\label{eqn:Proof_3}
\end{equation}

The characteristic polynomial of $\Matrix{\Phi}$ is:
\begin{equation}
\det\left( \lambda \Matrix{I} - \Matrix{\Phi} \right)
=
\det\left( \lambda \Matrix{I} - \left( \alpha \Matrix{I} + \Matrix{A}^\Transpose \right) \right)
=
(\lambda - \alpha)^n
.
\label{eqn:Proof_4}
\end{equation}

Comparing \refEq{eqn:Proof_4} with \refEq{eqn:CtrlDesign_K_PolePlacement} and \refEq{eqn:CtrlDesign_K_PolePlacement_Param} when $\alpha = \omega_\mathrm{CL}$ we find the characteristic polynomial to be:
\begin{equation}
\det\left( \lambda \Matrix{I} - \Matrix{\Phi} \right)
=
\lambda^n - \sum_{i = 1}^n (-1)^{(n-i)} \cdot k_{\mathrm{BW},i} \cdot \lambda^{i-1}
.
\label{eqn:Proof_5}
\end{equation}

This allows us to apply the Cayley--Hamilton theorem to \refEq{eqn:Proof_3}, with $\Matrix{\Phi}^n = \sum_{i = 1}^n (-1)^{(n-i)} \cdot \Matrix{\Phi}^{(i-1)} \cdot k_{\mathrm{BW},i}$ we finally obtain:
\begin{equation}
\Vector{p}_n
= \Vector{k}_\alpha
= \frac{1}{2} \Vector{k}_\mathrm{BW}
.
\label{eqn:Proof_Result}
\end{equation}

This concludes the proof.
As the analytical solution of the algebraic Riccati equation \refEq{eqn:Alpha_Riccati}, it provides a link between optimal control and pole placement for linear ADRC.
\qed
\end{pf}

\begin{rem}
Due to the duality of the design problem, a proof of the half-gain relation for the extended state observer design (with $k_\mathrm{ESO} \cdot \omega_\mathrm{CL} = \alpha$) can be constructed in the same manner.
\end{rem}


\section{Examples}
\label{sec:examples}

Aim of this section is to provide visual insights into an ADRC-based control loop when using half-gain tuning for the controller, the extended state observer, or both.
For this purpose we can restrict ourselves to a second-order plant with normalized gain and eigenfrequency:
\begin{equation}
P(s) = \frac{1}{s^2 + 2 s + 1}
.
\label{eqn:Example_Plant}
\end{equation}

Since ADRC is almost insensitive to the damping ratio, especially of underdamped systems, cf.\ \cite{Herbst:2013}, the informative value of our example will not be compromised by the particular choice of critical damping in $P(s)$.

Bandwidth parameterization is applied to a second-order ADRC ($n = 2$) using $\omega_\mathrm{CL} = 1\,\mathrm{rad/s}$, $k_\mathrm{ESO} = 10$, and $b_0 = 1$. Four cases are being compared: (1) unmodified bandwidth tuning, (2) applying half-gain tuning only to the outer control loop (``K/2 controller''), (3) applying half-gain tuning only to the ESO (``L/2 observer''), and (4) half-gain tuning for both controller and observer.


\subsection{Impact on Open-Loop Characteristics}
\label{sec:examples_openloop}

\begin{figure}[ht]
    \centering%
    \includegraphics{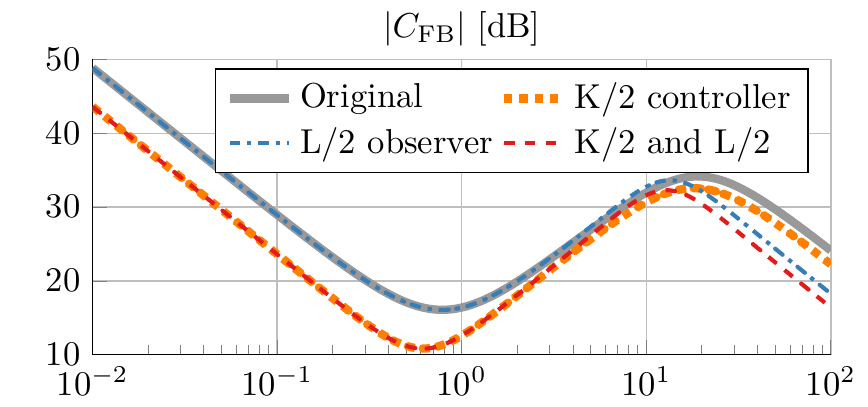}%
    \\%
    \includegraphics{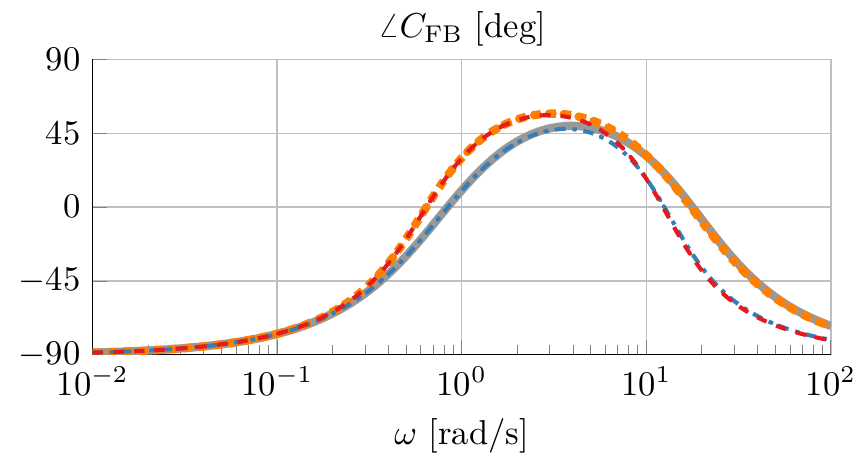}%
    \caption{Comparison of the feedback controller transfer functions with or without half-gain tuning.}
    \label{fig:Comparison_CFB}
\end{figure}

\begin{figure}[ht]
    \centering%
    \includegraphics{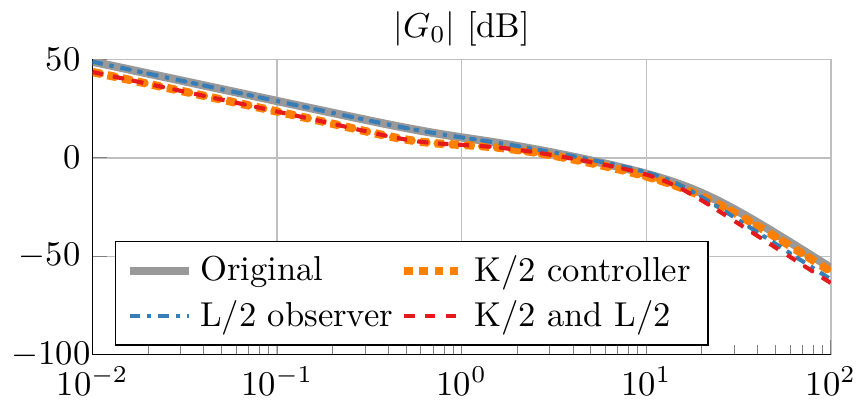}%
    \\%
    \includegraphics{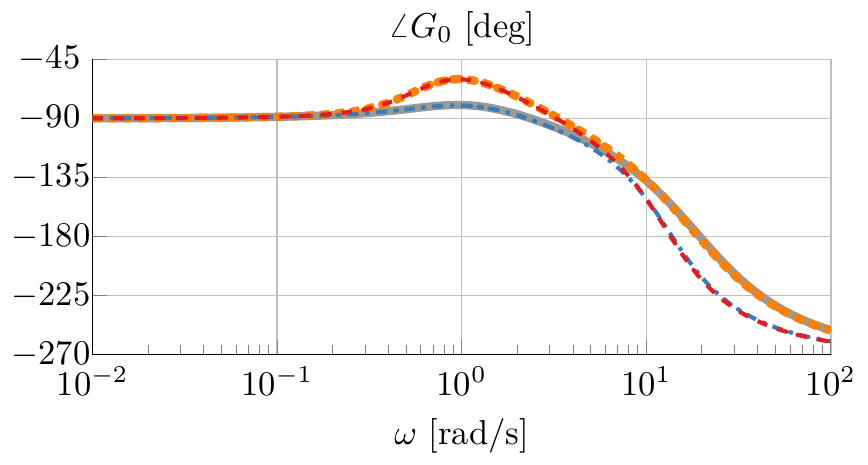}%
    \caption{Comparison of the open-loop gain transfer function with or without half-gain tuning.}
    \label{fig:Comparison_G0}
\end{figure}

For stability and dynamics, the feedback controller part of an ADRC control loop is essential. In \refFig{fig:Comparison_CFB} the transfer functions from controlled variable $y$ to control signal $u$ are compared for the four possible cases. Additionally, the loop gain transfer functions are being compared in \refFig{fig:Comparison_G0}.

The most interesting result might be that half-gain observer tuning (``L/2'' case) provides significantly improved high-frequency damping while having almost no impact on the lower frequencies up to and including the crossover frequency. On the other hand one has to expect some low-frequency performance penalty when (additionally or solely) applying half-gain controller tuning (``K/2'' cases).


\subsection{Impact on Closed-Loop Characteristics}
\label{sec:examples_closedloop}

\begin{figure*}[p]
    \centering%
    \includegraphics{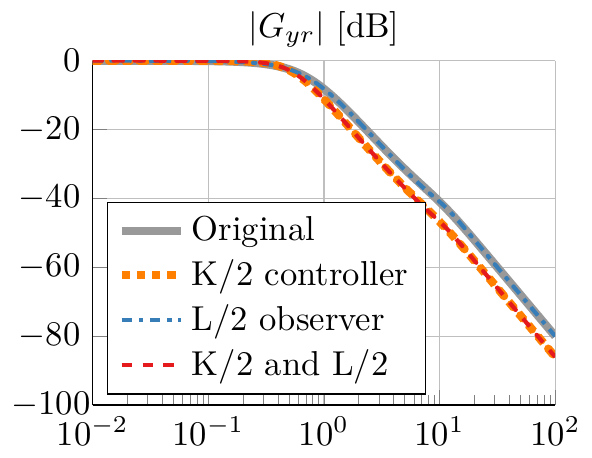}%
    \includegraphics{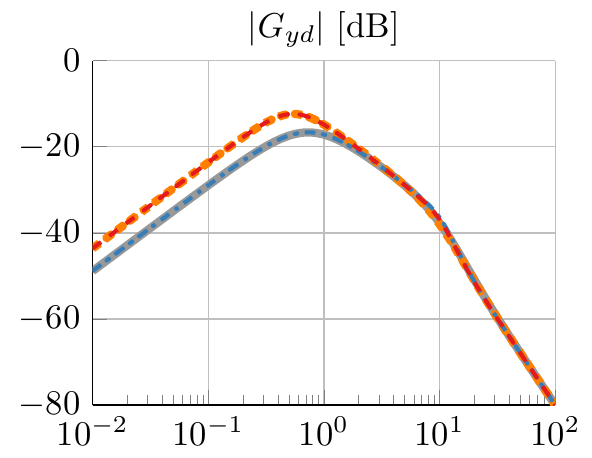}%
    \includegraphics{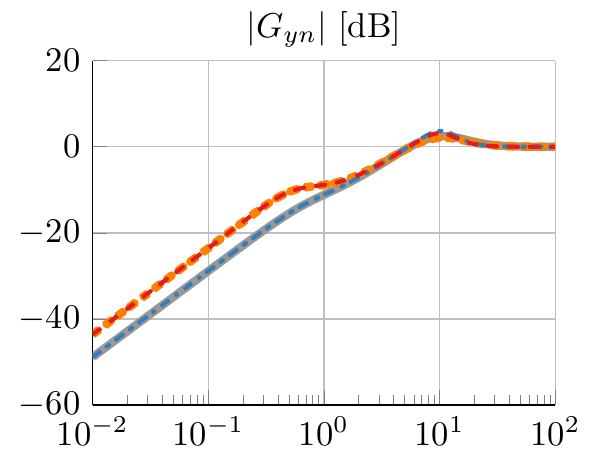}%
    \\%
    \includegraphics{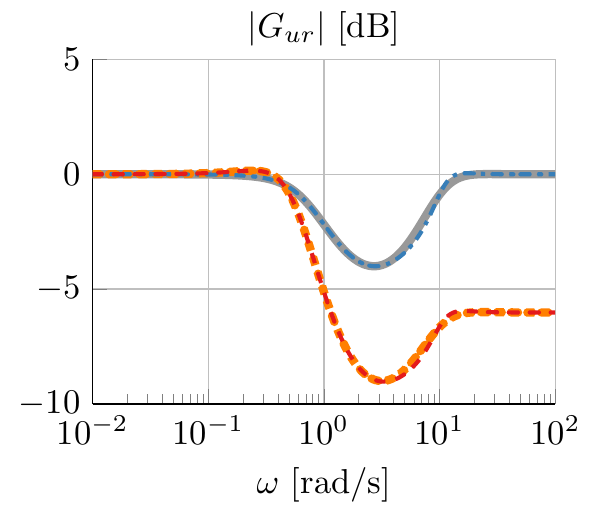}%
    \includegraphics{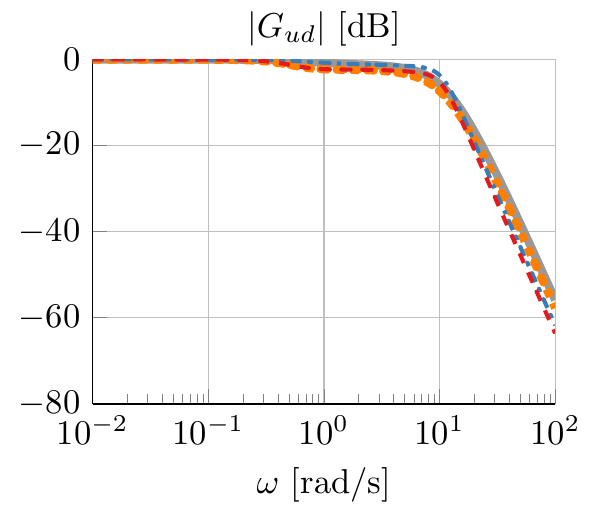}%
    \includegraphics{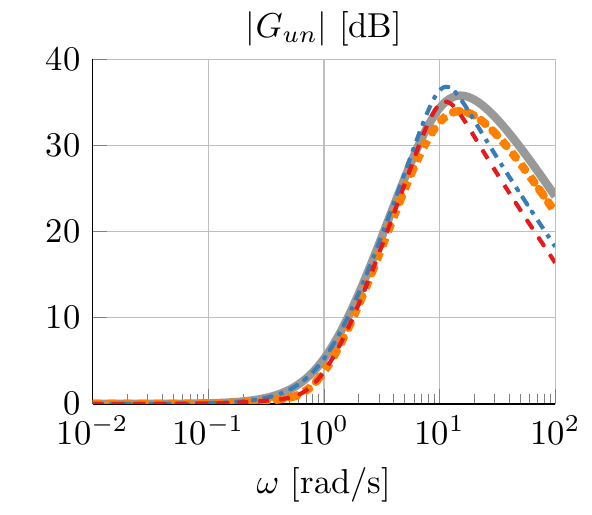}%
    \caption{Gang-of-six comparison (frequency domain) with or without half-gain tuning for controller and/or observer within ADRC. As predicted in \refSec{sec:examples_openloop}, the half-gain observer (``L/2'') case provides enhanced high-frequency damping in $G_{un}(\mathrm{j}\omega)$ almost without any side-effects on other performance criteria. The ``K/2'' cases, on the other hand, will---while still yielding some additional high-frequency damping in $G_{un}$---involve slower reaction to reference signal changes and more overshoot induced by disturbances at the plant input.}
    \label{fig:Comparison_GangOfSix}
\end{figure*}

\begin{figure*}[p]
    \centering%
    \includegraphics{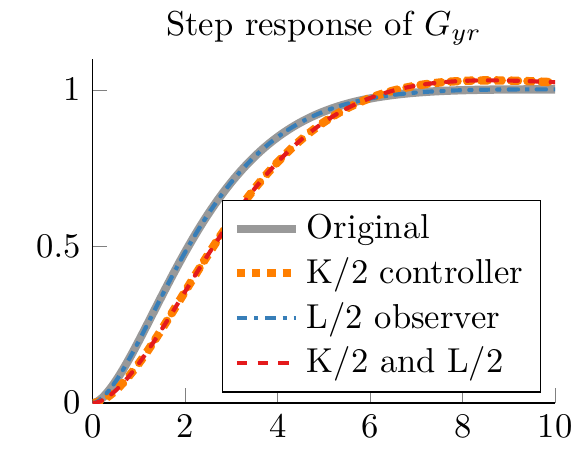}%
    \includegraphics{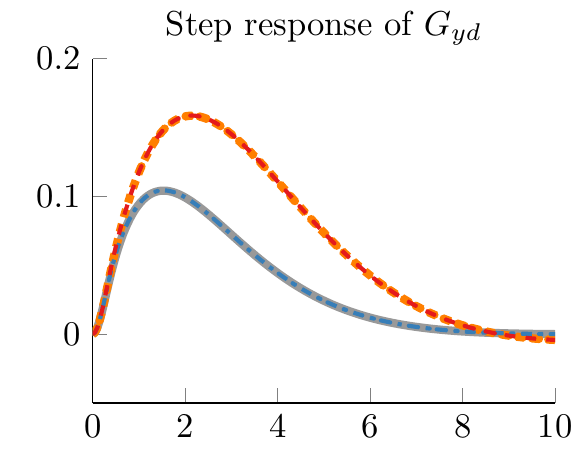}%
    \includegraphics{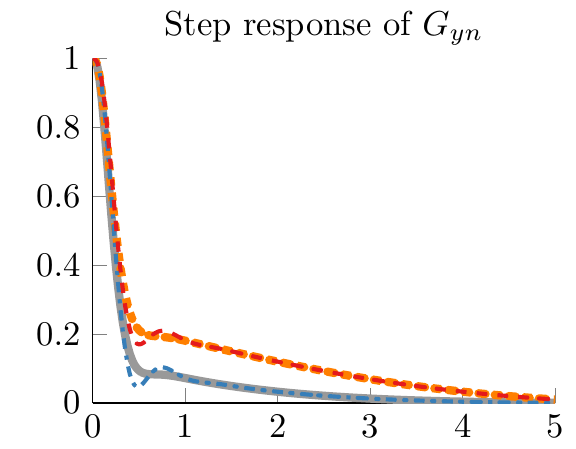}%
    \\%
    \includegraphics{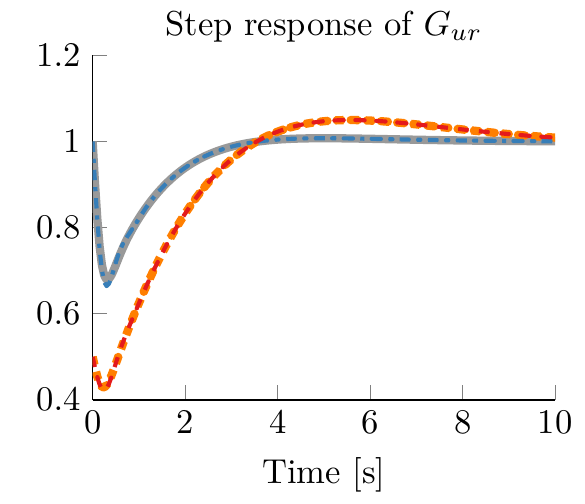}%
    \includegraphics{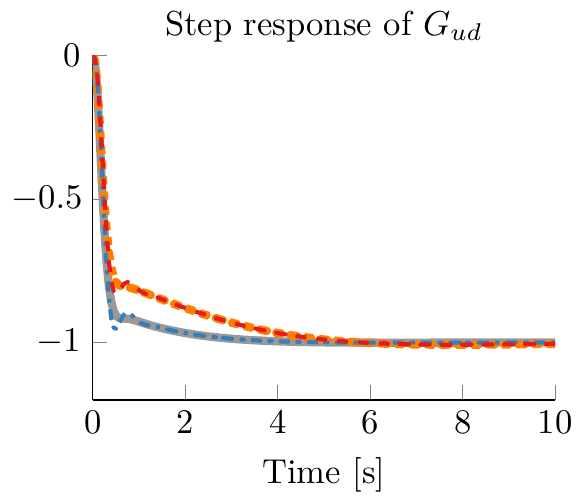}%
    \includegraphics{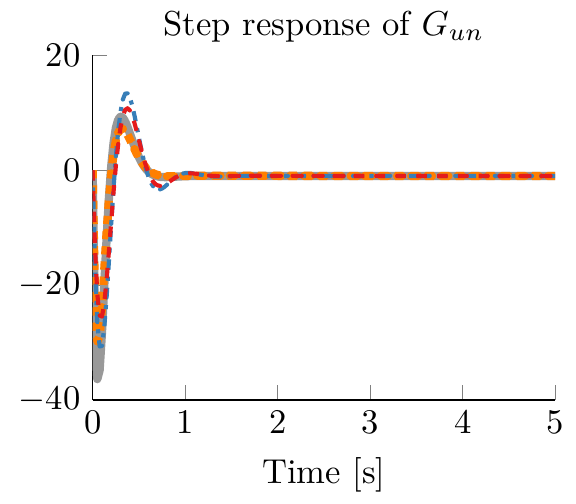}%
    \caption{To ease and support the interpretability of \refFig{fig:Comparison_GangOfSix}, a time-domain perspective is given in this figure with the step responses of the gang-of-six transfer functions with or without half-gain tuning.}
    \label{fig:Comparison_GangOfSix_Step}
\end{figure*}

With the control loop signals denoted as in \refFig{fig:ADRC_ControlLoop}, the ``gang of six'' transfer functions are defined as
\begin{equation}
\begin{pmatrix}
y(s)
\\
u(s)
\end{pmatrix}
=
\begin{pmatrix}
G_{yr}(s)  &  G_{yd}(s)  &  G_{yn}(s)
\\
G_{ur}(s)  &  G_{ud}(s)  &  G_{un}(s)
\end{pmatrix}
\cdot
\begin{pmatrix}
r(s)
\\
d(s)
\\
n(s)
\end{pmatrix}
,
\label{eqn:GangOfSix}
\end{equation}
providing frequency-domain insights for a two-degrees-of-freedom control loop as is the case with ADRC, cf.\ \cite{AstromMurray:2008}. For the four cases in our example, they are presented and discussed in \refFig{fig:Comparison_GangOfSix} and \refFig{fig:Comparison_GangOfSix_Step}.

While not shown here for brevity, a discrete-time implementation of ``K/2'' and ``L/2'' design was successfully tested as well, exhibiting the desired noise reduction in the control signal $u$.

\section{Conclusion}
\label{sec:conclusion}

A new ``half-gain tuning'' rule for linear active disturbance rejection control (ADRC) based on the so-called $\alpha$-controller design was introduced.
Compared to the common ``bandwidth parameterization'' approach, similar closed-loop dynamics can be achieved with lower (halved) feedback gains, therefore reducing the noise sensitivity of ADRC.

In view of the examples presented in \refSec{sec:examples}, a recommendation emerges to start with half-gain tuning for the observer. This has the least impact on the closed-loop dynamics compared to bandwidth parameterization, while already providing a significant reduction of control signal sensitivity to measurement noise.

While being the analytical solution of an algebraic Riccati equation, the proposed feedback gains can simply be obtained from a bandwidth parameterization design by halving the gains, as proved in this paper, establishing a link between pole placement and optimal control.


\begin{ack}
Gernot Herbst would like to thank Michael Buhl for drawing his attention to the $\alpha$-controller approach.
\end{ack}

\end{document}